\documentclass[vecphys]{svmult}

\usepackage{makeidx}         
\usepackage{graphicx}        
\usepackage{multicol}        
\usepackage[bottom]{footmisc}

\makeindex             

\begin{document}

\title*{A Discrete
  Representation of Einstein's Geometric Theory of Gravitation: The
  Fundamental Role of Dual Tessellations in Regge Calculus}
 \titlerunning{Voronoi-Delaunay Lattices in Regge Calculus} 
 
\author{Jonathan R. McDonald \and
Warner A. Miller}
\authorrunning{Miller \& McDonald} 
\institute{Department of Physics, Florida Atlantic University, Boca
  Raton, FL 33431, USA
\texttt{wam@physics.fau.edu}}
%
%
\maketitle

In 1961 Tullio Regge provided us with a beautiful lattice
representation of Einstein's geometric theory of gravity.  This Regge
Calculus (RC) is strikingly different from the more usual finite
difference and finite element discretizations of gravity.  In RC the
fundamental principles of General Relativity are applied directly to a
tessellated spacetime geometry.  In this manuscript, and in the spirit
of this conference, we reexamine the foundations of RC and emphasize
the central role that the Voronoi and Delaunay lattices play in this
discrete theory.  In particular we describe, for the first time, a
geometric construction of the scalar curvature invariant at a vertex.
This derivation makes use of a new fundamental lattice cell built from
elements inherited from both the simplicial (Delaunay) spacetime and
its circumcentric dual (Voronoi) lattice. The orthogonality properties
between these two lattices yield an expression for the vertex-based
scalar curvature which is strikingly similar to the corresponding and
more familiar hinge-based expression in RC (deficit angle per unit
Voronoi dual area).  In particular, we show that the scalar curvature
is simply a vertex-based weighted average of deficits per weighted
average of dual areas. What is most striking to us is how naturally
spacetime is represented by Voronoi and Delaunay structures and that
the laws of gravity appear to be encoded locally on the lattice
spacetime with less complexity than in the continuum, yet the
continuum is recovered by convergence in mean. Perhaps these prominent
features may enable us to transcend the details of any particular
discrete model gravitation and yield clues to help us discover how we
may begin to quantize this fundamental interaction.

\section{Why Regge Calculus?}
\label{sec1}

If nature is, at its very foundation, best expressed as a discrete
theory -- a finite representation based upon the elementary quantum
phenomena, then how will such a corpuscular structure reveal itself by
observation or measurement?  If the quantum character of nature has
taught us anything, it has taught us that (1) the observer and the
observed are demonstrably coupled, and that (2) the basis of
observation lay on the distinguishability between complementary
observables. The currency of the quantum is information -- information
communicated from observer to observer in the form of quantum bits
(qbits.)  It is hard to imagine setting the stage for such a theory,
and many have worked to do so. We believe that one small step can be
made in this direction by examining a discrete representation of
Einstein's 1915 classical theory of gravitation -- one of the purest
geometric theories of nature we know. One may hope that by studying a
discrete representation of gravitation, one may be able to glean some
of the fundamental features of the discretization that may yield way
points to true understanding of the basic building blocks of nature.

Fortunately, in 1961 Tullio Regge introduced a beautiful simplicial
representation of Einstein's geometric theory of gravitation
\cite{R61}. It is referred to in the literature as Regge Calculus
(RC).  This approach represents the curved 4-dimensional spacetime
geometry, which is so central to General Relativity (GR), as a
discrete 4-dimensional simplicial lattice.  The interior geometry of
each simplex in this lattice is that of the flat spacetime geometry of
Minkowski space.  Since the beginnings of RC many researchers have
successfully applied RC to many problems in both classical and quantum
GR (references in \cite{TW92}).

What is so striking to us is how naturally the concepts of the Voronoi
and Delaunay lattices \cite{book} appear in every facet of RC, and we
think this representation applies to any discretization of spacetime.
Perhaps this is not so surprising as witnessed by these proceedings of
the vast applications the concepts of these lattices have throughout
nature. We see them naturally appearing in nature from biological
structures to anthropological applications, and from cosmology to the
quantum and foams.  It is the purpose of this paper to demonstrate
concretely the necessary and natural role these Voronoi and Delaunay
structures have in discrete models of gravity. \cite{CMS82, CMS84,
  H86, M86, M92, L82} That these two circumcentric lattices have
encoded in them beautiful orthogonality properties, natural duality as
well as a democratic way of assigning a unique and indisputable
``territory'' to every single event (atom, vertex, star, observer
etc...) gives us hope as to their universality -- hope that they can
be used, in part, as an austere description of fundamental physics.

After briefly introducing some of the key features of Einstein's
theory of GR (Sec.~\ref{sec2}) that we use in RC, we will describe how
to understand the curvature of a lattice spacetime in in RC
(Sec.~\ref{sec3}). A detailed analysis of this lattice curvature, we
show, yields a natural tiling of the lattice spacetime in terms of a
new and fundamental 4-dimensional lattice cell -- a hybridization of
both the Voronoi lattice and the Delaunay lattice
(Sec.~\ref{sec4}). We demonstrate the usefulness of this new spacetime
building block in RC by re-deriving the RC version of the
Einstein-Hilbert action principle, and in the local construction of
the Einstein tensor in RC. This coupling of the Voronoi and Delaunay
lattices is further used to provide the first geometric construction
of the scalar curvature invariant of which we are aware
(Sec.~\ref{sec5}-\ref{sec6}).  In particular, we show that the scalar
curvature is simply a vertex-based weighted average of deficits per
weighted average of dual areas.  We provide (Sec.~\ref{sec7}) an
example of this vertex-based scalar curvature and discuss its
convergence-in-mean to the continuum counterpart.

\section{Gravitation and the Curvature of Spacetime}
\label{sec2}

The geometric nature of the gravitational interaction is captured in
Einstein's theory of GR  by a single equation \cite{MTW73}.
\begin{equation}
\label{EEQ}
\underbrace{G_{\mu\nu}}_{\begin{array}{c} Curved \\ Spacetime \\ Geometry\end{array}} = 
8 \pi \underbrace{T_{\mu \nu}}_{\begin{array}{c} Matter \\ and \\ Fields\end{array}}
\end{equation}
Here the curvature of spacetime, as encapsulated by the ten-component
Einstein tensor $G_{\mu\nu}$, acts on matter telling it how to curve,
and in response these non-gravitational matter sources (the
stress-energy tensor $T_{\mu\nu}$) react back on spacetime telling it
how to curve. Central to this theory of gravity is the 4-dimensional
curved spacetime geometry.  This geometry is represented by a
$4\times4$ metric tensor $g_{\mu\nu}$, or by the edge lengths of a
tessellated geometry as we will see below. 

\begin{figure}
\centering
\includegraphics[width=0.5\linewidth]{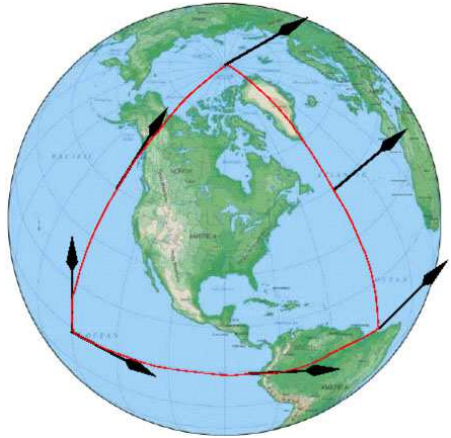}
\caption{{\it The curvature of the Earth, idealized though it is as a
    sphere of radius $R_{\oplus}$, as revealed by parallel transport
    of a vector around a closed loop:} We can demonstrate this
  equation by application to the 2-dimensional surface of the Earth.
  We show in this figure the clockwise circumnavigation of one
  quadrant of the northern hemisphere of the Earth. The starting point
  is the northward pointing vector on the equator, and three snapshots
  of this vector as it is parallel transported to the North Pole. The
  parallel transport of the vector is with respect to the geometry
  intrinsic to the surface of the sphere as there is no need
  whatsoever of relying on the embedding of the sphere in the
  Euclidean 3-dimensional geometry. That this vector is parallel
  transported parallel to itself is demonstrated that at every point
  on its northward journey up the longitudinal direction its always
  tangent to its path. Upon reaching the north pole, we take great
  care not to take the energy needed to rotate the vector even though
  our new path takes an abrupt turn to the right as we head southward
  down another longitude. In the beginning of this southward segment
  of our journey, at the North Pole, the vector is normal to this new
  longitudinal direction, and remains so because it is transported
  parallel to itself all the way back to the equator. At the equator
  the path takes another abrupt right angle turn toward the west. The
  vector which has never been rotated remains pointing eastward on the
  equator as it is parallel transported back along the equator to its
  starting position. As the vector returns to its starting point one
  can't help but notice that it is rotated perpendicular to its
  original northward direction, although the vector was never rotated
  throughout its entire journey -- a journey which swept out an area
  $1/8$ of the surface of the Earth. The reason for the net rotation
  is clear.  The path circumnavigated contained a geometry of constant
  curvature.  It is through this process of parallel transport that we
  can most clearly understand and analyze the curvature of a lattice
  spacetime geometry in RC.  }
\label{fig:earth}
\end{figure}

Central to Einstein's theory is curvature and key to our description
of RC in (Sec.~\ref{sec3}) is the construction of the Riemann curvature
tensor and the scalar curvature invariant on a Voronoi/Delaunay
spacetime lattice. The components of the curvature tensor can be
understood geometrically by either geodesic deviation or by the
parallel transport of a vector around a closed loop.  In RC is is most
convenient to implement the latter. Here we illustrate curvature by a
familiar example, the Gaussian curvature by way of parallel transport
of a vector around a closed loop (Fig~\ref{fig:earth}).
\begin{equation}\label{GaussianCurv}
\mbox{Curvature} = K =
\frac{ \left( \begin{array}{c} \mbox{Angle Vector} \\ \mbox{Rotates}\end{array} \right) }
     { \left( \begin{array}{c} \mbox{Area} \\ \mbox{Circumnavigated}\end{array} \right)} = 
\frac{\frac{\pi}{2}}{\frac{1}{8} \left( 4\pi R_{\oplus}^2\right)} = 
\frac{1}{R^2_{\oplus}}
\end{equation} 

In the 2-dimensional example of curvature shown above, only a single
component of curvature is needed to completely characterize the
curvature.  There is only one tangent plane at each point on the
surface that can be used to form a loop to circumnavigate.  However
in 4-dimensions there are many more orientations of areas to
circumnavigate (Table~\ref{table:xyzt}). In addition to a scalar
curvature, we actually need the structure of a tensor to characterize
the curvature. This curvature is represented by the Riemann curvature
tensor (\ref{rotop}). Each of its components (twenty in the four
dimensions of spacetime) are as easily described in the same
fundamental way in which we obtained the curvature of the earth in
Fig.~\ref{fig:earth} -- by parallel transport of a vector about a
vanishingly small closed loop.
\begin{equation}
\label{rotop}
{\cal R} = 
\left(\begin{array}{c}\mbox{Rotation}\\\mbox{Operator}\end{array}\right) =
- \overbrace{e_{\alpha} \wedge e_{\beta}}^{\begin{array}{c} Rotation \\  Bivector\end{array}}
R^{|\alpha \beta |}{}_{| \mu \nu|}  
\underbrace{dx^\mu \wedge dx^\nu}_{\begin{array}{c} Orentation \\  of\  Area \\ Circumnavigated\end{array}}
\end{equation}
This rotation operator acts on an oriented area (area bivector or
$A^{\mu\nu}\, e_\mu \wedge e_\nu$) and returns a rotation bivector
(the exterior product of the original vector with the vector parallel
transported around the area back to its starting point).
\begin{table}[t]
\label{table:xyzt}
\centering
\begin{tabular}{c c c c c c c}
\hline\hline 
Dimension &    &       &      &     &      &     \\ [0.5ex]
\hline
    2-D: & x-y &       &      &      &     &     \\
    3-D: & x-y &  x-z  & y-z  &      &     &     \\
    4-D: & t-x &  t-y  & t-z  & x-y  & x-z & y-z \\ [1ex]
\hline
\end{tabular}
\caption{
Identification of the planes of rotation for various dimensions. The number 
of independent Riemann tensor components increases with dimension, in two 
dimensions the rotation operator is completely described by a single 
component, in 3-dimensions by six and in the 4-dimensions by the
20~independent components of the Riemann curvature tensor.
}
\end{table}
The twenty components of the Riemann curvature tensor completely
describe the curvature of the spacetime. From it we can construct the
scalar curvature invariant by tracing over its components.
\begin{equation}
\left(
\begin{array}{c}\mbox{Scalar Curvature}\\\mbox{Invariant}\end{array}  
\right) = R = R^{\mu\nu}{}_{\mu\nu}.
\end{equation}

Hilbert showed that this scalar curvature invariant plays a central
role into the inner workings of GR when he introduced the
Einstein-Hilbert action principle. It embodies all of Einstein's theory
in a single variational principle, and the quantity varied is the
scalar curvature times the proper 4-dimensional volume element of
spacetime.
\begin{equation}
\label{EH}
I = \frac{1}{16\pi} \int {}^{(4)}R\ d^{(4)}V_{proper}
\end{equation}
\begin{equation}
\label{EH2}
\underbrace{\delta I\  =\  0}_{g_{\mu\nu} \longrightarrow g_{\mu\nu} + \delta g_{\mu\nu}}
\end{equation}

This action principle, when added to an appropriate action for the
non-gravitational sources, yields Einstein's field equations
(\ref{EEQ}).

In the remaining sections we will demonstrate that the Voronoi and
Delaunay lattices play a fundamental role in the definition of
Einsteins theory on a lattice spacetime. Each of these quantities (1)
the Riemann curvature tensor, (2) the scalar curvature invariant, and
(3) the Einstein equations, can be constructed on a lattice spacetime
geometry. Each of these three covariant objects rely equally on the
Voronoi lattice as well as its dual simplicial Delaunay lattice for
their definition.

\section{Curvature in Regge Calculus}
\label{sec3}

\begin{figure}
\centering
\includegraphics[width=0.5\linewidth]{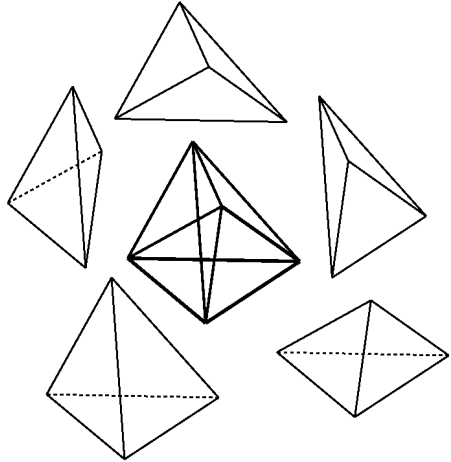}
\caption{{\em A 4-Dimensional Simplex with its Tetrahedral Boundary:}
  A 4-dimensional simplex is shown in the central region of this
  illustration. The simplex consists of 5 vertices, 10 edges, 10
  triangles and 5 tetrahedrons.  We explicitly show each of the
  simplex's five bounding tetrahedrons exploded around its perimeter.}
\label{fig:simplex}
\end{figure}

\begin{figure}
\centering
\includegraphics[width=0.6\linewidth]{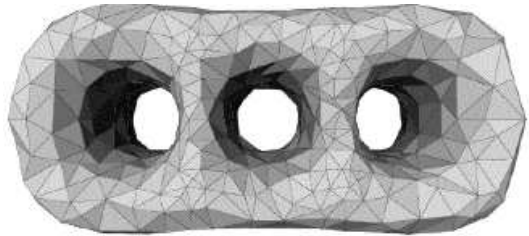}
\caption{{\it A Triangulated 2-surface of Complex Topology: }One of
  the strengths of lattice gravity is its ability to model complex
  topologies.  We use this figure to illustrate a typical
  2-dimensional triangulated surface with complex topology.  While it
  is easiest to depict two-dimensional curved surfaces here, our
  interests lie primarily in the description of spacetime.  This
  requires the transition from tessellations with triangles to
  tessellations with 4-simplices. }
\label{fig:lattice}
\end{figure}

In RC the spacetime geometry is represented by a 4-dimensional Voronoi
lattice, or its circumcentric dual or Delaunay lattice.  Each block in
this Delaunay lattice is a simplex, i.e. a 4-dimensional triangle
(Fig.~\ref{fig:simplex}). Each simplex in this discrete spacetime
lattice has the geometry of flat Minkowski spacetime. The geometry of
each simplex is determined by its ten edges. This simplex ia the tile
for the latttice spacetime we use in RC to describe the gravitational
field.  This tessellated spacetime geometry can be visualized as the
4-dimensional analogue of the familiar architectural geodesic domes.
The geodesic domes or other more homotopically interesting
triangulated surfaces are often constructed by a lattice of triangles,
2-dimensional simplexes (Fig.~\ref{fig:lattice}).  In higher
dimensions (d), which is virtually impossible to depict in a single
illustration, the tiles are d-dimensional simplexes.

\begin{figure}
\centering
\includegraphics[width=0.5\linewidth]{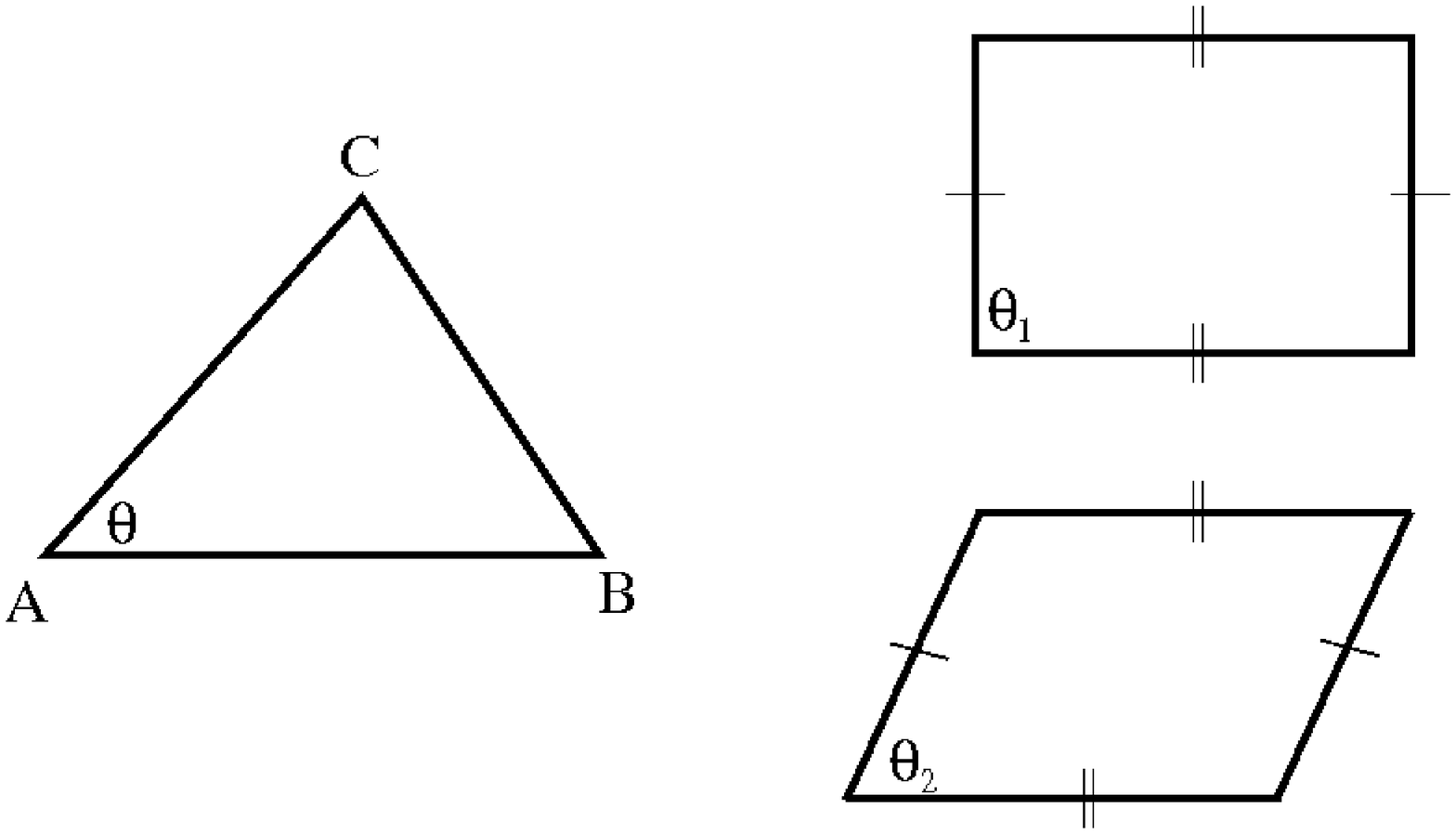}
\caption{{\it Rigidity of simplexes and their use in RC :} (Left) A
  typical triangle $ABC$ in a 2-dimensional  Delaunay lattice.  One
  need only specify the lengths of its three   edges, $\overline{AB}$,
  $\overline{AC}$ and $\overline{BC}$, to uniquely determine the
  geometry of the triangle (simplex). In particular, the area of the
  triangle $\triangle_{ABC}$ and its internal angles $\sin{\theta} = 2
  \triangle_{ABC}/  \overline{AB}\,\overline{AC}$ are explicit
  functions of the length of its three edges. In this sense the
  2-simplex is rigid. (Right) Even though the four edges of the rectangle
  (upper right) are specified, its internal angle $\theta_1$ is not
  uniquely determined.  The rectangle can be collapsed to a
  parallelogram (lower right) while preserving the lengths of each of
  its four edges. This non-simplicial block is not rigid and
  ${\theta}_1 \neq {\theta_2}$.}
\label{fig:rigid}
\end{figure}

Simplicies are often used in RC because the geometry of
each simplex can be uniquely determined by the lengths of its ten
edges.  It is therefore convenient to represent the spacetime geometry
by a simplicial Delaunay lattice. Each polytope in the Delaunay
lattice is a simplex and are therefore rigid (Fig.~\ref{fig:rigid}).
This lattice and its dual Voronoi lattice form the basis for RC. In
the simplicial lattice, the length ($L$) of an edge of a simplex is
understood  as the proper distance/time between its two bounding
vertices,  and it represents the RC analog of the metric components.
\begin{equation}
L^2 \ \Longleftrightarrow \ \left( g_{\mu\nu} dx^{\mu} dx^{\nu} \right)
\end{equation}
Correspondingly, timelike edges will have a negative square magnitude while
square of the length of a spacelike edge will be positive. 

In RC the curvature of a Delaunay lattice is concentrated at its
co-dimension two elements. We will refer to these foci of curvature as
hinges, $h$. The hinges in a 2-dimensional lattice are the vertices,
in three dimensions the components of the Riemann curvature tensor are
concentrated along the edges, and in the four dimensions of spacetime
the triangles are the hinges. In the remainder of this section we
examine the construction of the components of the Riemann curvature
tensor on simplicial lattices first in 2-dimensions, then in
3-dimensions and finally in 4-dimensions. We demonstrate that the
Voronoi areas dual to the hinges provide a natural measure for the
curvature.

In 2-dimensions the building blocks are triangles. The spacetime
geometry interior to each triangle is flat. Curvature, when present,
is only concentrated at each vertex (co-dimension 2). The curvature
represents a conic singularity.  In other words, the net rotation of a
vector when transported parallel to itself around a vertex hinge
remains unchanged as one shrinks the closed loop.  This constant rotation angle 
is referred to as the deficit angle ($\epsilon_h$).

\begin{figure}
\centering
\includegraphics[width=0.8\linewidth]{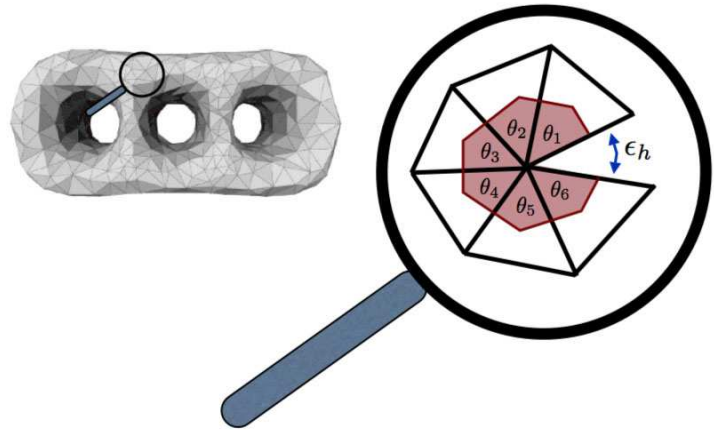}
\caption{{\it Curvature is concentrated at the vertex hinges in two
    dimensions:} (Left) The 2-dimensional triangular tessellation
  approximates triple torus. Here we examine the curvature
  concentrated at a particular vertex of this triple torus. (Right) When we
  magnify the vertex, we find it is the meeting place of six
  triangles, and that these six triangles form a convex hull. To
  flatten these six triangles onto a Euclidean plane the hull need to
  split along one of its edges.  This angle of split, or deficit
  angle, captures the conic curvature.  The deficit angle associated with a 
  vertex is revealed as the sum of the internal angles' difference
  from $2\pi$. The Voronoi area dual to the vertex (the shaded area) gives a
  natural closed loop for the definition of the Gaussian curvature.}
\label{fig:2D}
\end{figure}

\begin{equation}\label{defangle}
  \underbrace{\epsilon_h = 2\pi -\sum_{i=1}^{6}
    \theta_i}_{\begin{array}{c} Deficit\, Angle  \\  Angle\,Vector\,Rotates\end{array}}
\end{equation} 
Here $\theta_i$ is the internal angle of the $i'th$ triangle at vertex
hinge $h$ as illustrated in Fig.~\ref{fig:2D}. In order to calculate
the curvature, an appropriate closed loop surrounding our vertex hinge
must be chosen for parallel transport.  Since the hinge represents a
conic singularity, the Gaussian curvature is independent of the area
of the loop enclosing the vertex; however, the Voronoi area
dual to the vertex hinge represents a natural choice for this area.  In
2-dimensions, this area gives a natural closed loop that does not
overlap any other closed loops on the tessellated
geometry. Furthermore, it represents the set of points on the
simplicial lattice close to this vertex than to any other vertex.  In
this sense it provides a democratic measure of the area assigned to
each hinge.  Using the deficit angle and the dual Voronoi area, the curvature
(\ref{GaussianCurv}) takes the usual form.   

\begin{equation}\label{2Dcurvature}
K_h = \frac{\epsilon_h}{A^*_h} 
\end{equation}

\begin{figure}
\centering
\includegraphics[width=0.6\linewidth]{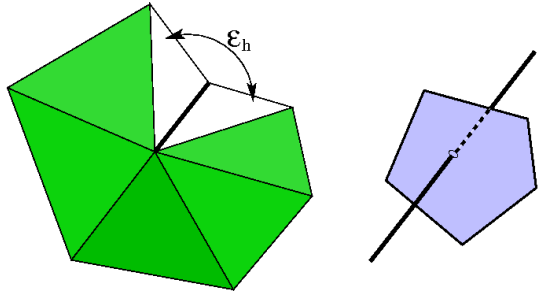}
\caption{{\it Hinges in 3-dimensions:  }In 3-dimensions, curvature is
  concentrated along edges. (Left) The   deficit angle, the deficit
  from 2$\pi$ in the  sum of dihedral angles on the hinge, determines
  the  angle by which a vector carried around a loop surronding the
  hinge comes back rotated.  (Right) The dual to the edge is
a 2-dimensional face of the Voronoi diagram associated with the
triangulation. This area serves as a natural loop for determination of
the curvature associated with the hinge.}
\label{fig:3D}
\end{figure}

In 3-dimensions the building blocks are tetrahedra.  The
interior of each of these  base building blocks is flat, and
curvature is concentrated on the co-dimension 2 edges of the 
tessellationon the edges of the simplicial lattice.  As one transports
a vector around a loop surrounding the edge, the vector ordinarily comes back
rotated by an angle $\epsilon_{h}$ which is still given
by~(\ref{defangle}) and shown in Fig.~\ref{fig:3D}. However, the $\theta_{i}$'s now
represent the dihedral angles of the $i$'th tetrahedron formed between
its two  triangular faces that share the hinge. Dual to the edges are
the 2-dimensional Voronoi areas, which will again play the role of
loops for parallel transport as they still represent a proper tiling
of the spacetime, as well as a natural weighting for each hinge.  The
curvature $^{(3)}K_h$ takes the same form as in 2-dimensions and is shown below. 

\begin{equation}
^{(3)}K_h = \frac{\epsilon_h}{A^*_h}
\end{equation}

The story remains similar in 4-dimensions.  The 4-simplex of
Fig.~\ref{fig:simplex} takes over the role of Delaunay cells. The
interior of the 4-simplices are flat spacetime, and curvature is then 
concentrated on the triangular faces of the 4-simplices.  Curvature is
now measured by parallel transporting a vector around the triangular
faces of the Delaunay lattice, which ordinarily will come back rotated
by the deficit angle $\epsilon_h$.  This deficit angle remains of the
form of~(\ref{defangle}) where the $\theta_{i}$'s are now interpreted
as the hyper-dihedral angles between tetrahedral faces of the $i$'th
4-simplex sharing the triangular hinge. Dual to the triangular hinges
are still the 2-dimensional Voronoi faces, which again provide a
natural closed loop for parallel transport.  Taking the deficit angle
and the Voronoi area dual to the hinge, the curvature $^{(4)}K_h$
maintains its general character from the 2- and 3-dimensional forms.

\begin{figure}
\centering
\includegraphics[width=0.4\linewidth]{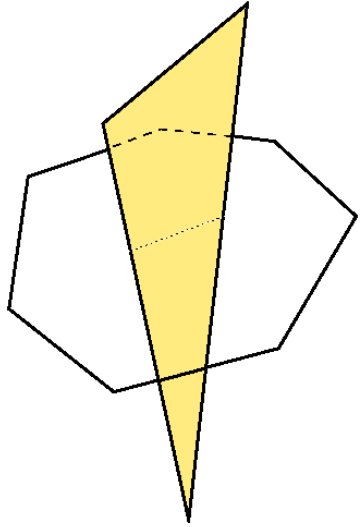}
\caption{{\it Hinges in 4-dimensions:  }In 4-dimensions, curvature is
  concentrated on co-dimension 2
triangular faces of the Delaunay lattice (shaded triangle).  Dual to hinges are the
2-dimensional Voronoi faces (the heptagon in this example) which
provide another natural area of closed parallel transport.}
\label{fig:4D}
\end{figure}

\begin{equation}
^{(4)}K_h = \frac{\epsilon_h}{A^*_h}
\end{equation}

Curvature in RC gives an example of the intertwined role that the Voronoi
and Delaunay tessellations play in this discrete theory of
spacetime.  The Voronoi areas provide a natural loop for parallel
transport while triangles of the Delaunay tessellation
provide a framework for the definition of the angle rotated by a
vector under parallel transport around the Voronoi area.  We will see in the
coming sections how the Voronoi and Delaunay lattices can be used to
define curvature in a way that is compatible with observation and measurment.

\section{A Fundamental Block Coupling the Voronoi and Delaunay
  Lattices Together}
\label{sec4}

\begin{figure}
\centering
\includegraphics[width=0.4\linewidth]{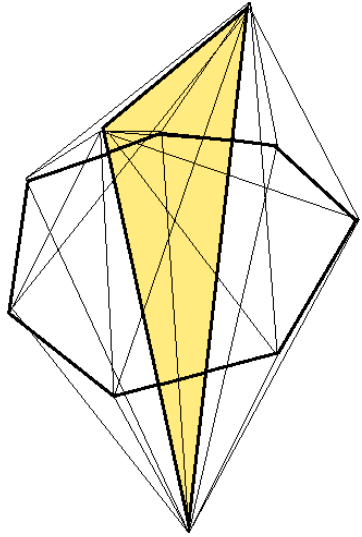}
\caption{{\it The Hybrid Block:} This definition of the hybrid block
  in RC forms a lattice of 4-dimensional blocks that incorporates many
  of the attractive features of the Delaunay and Voronoi lattices.
  These blocks naturally tile the spacetime and obtain rigidity from
  the underlying Delaunay lattice.  Interpretations of these hybrid of
  blocks are ripe with analogs from solid state physics. The Delaunay
  lattice serves as a spacetime analog of the Bravais lattice so often
  found in the theory of solids.  With this identification, the
  Voronoi cells might be considered analogs to the Brillouin zones for
  a non-periodic lattice.  The hybrid block becomes a hybrid between a
  spacetime lattice and a fundamental lattice structure of momentum
  space.  In anticipating future links with quantum theory, this leads
  us to consider this hybrid block as a precursor to a Wigner function
  on spacetime as it is precisely the Wigner function in standard
  quantum theory that attempts to link complementary variables into a
  qausi-probability function.  In a similar manner, the hybrid block
  forms a unique lattice composed of complimentary lattices of
  spacetime.}
\label{fig:FB}
\end{figure}

In the last section we emphasize that the curvature in a d-dimensional
simplicial spacetime takes the form of a conic singularity at each
point of the (d-2)-dimensional hinges (h). We noted that the
deflection of a ``d-axis gyroscope'' when carried one complete circuit
around a closed loop encircling a given hinge is indepedent of the
area of the loop.  We overcame this delimma in defining the curvature
of the hinge by introducing the Voronoi area ($A_h^*$) dual to the
hinge h  as a  natural and democratic measure for the distribution of this
curvature. If there are $n$ d-dimensional simplexes sharing hinge h,
then the Voronoi area ($A_h^*$) will be an $n$-gon (the heptagon in
Fig.~\ref{fig:4D}). Each corner of $h^*$ lies at the circumcenter of
one of the d-simplexes in the Delaunay geometry which share h.  We can
connect the $n$ vertices of $h^*$ to the triangle to form a
bi-pyramidal d-dimensional polytope.  This forms a new fundamental
lattice cell at each hinge h -- a hybrid polytope built from elements
from both the Voronoi ($h^*$) and Delaunay (h) hinges as shown
in Fig.~\ref{fig:FB}. This hybrid cell provides a proper tiling of the
spacetime.  Each hybrid cell having an $n$-gon $h^*$ can be
deconstructed into $n$ d-simpexes, where each of these simplexes is
formed by connecting an edge of $h^*$ with the (d-2) vertices of the
co-dimension 2 simplicial hinge h. In this sense, they are rigid
polytopes.

By construction, the Voronoi polygon $h^*$ is orthogonal to the
hinge h.  We use this orthogonality to show that the d-volume of these
hybrid cells are simply proportional to the product of the d-area of
the hinge ($A_h$) and the area ($A_h^*$) of its dual polygon $h^*$
\cite{M97}.  

\begin{equation}
\label{FBlk}
d^{(d)} V_{proper} = \frac{2}{d(d-1)} A_h A^*_h
\end{equation}

This is a co-dimension 2 version of the usual formula for the area of
a triangle as $\frac{1}{2} base \times altitude$.  These dual hybrid
Voronoi-Delaunay volume elements provide us with a natural
decomposition for our Hilbert action as well as a natural structure to
define the scalar curvature of any hinge, h.

The component of the Riemann curvature of this fundamental block is
now naturally expressed in terms of its intrinsic geometry. It not
only provides the orientation of the circumnavigated area ($A_h^*$),
it also yields the curvature hinge and its orientation ($A_h$).
\begin{equation}
\label{CExp}
^{(d)}R_h = d(d-1) ^{(d)}K_h = \frac{d(d-1)}{A^*_h} \epsilon_h
\end{equation}
Four features of this representation of curvature at a hinge in Regge
calculus are brought to the fore.  First, the curvature is
concentrated on the co-dimension 2 hinge, $h$. Second, the plane of
rotation is perpendicular to the the hinge, $h$, which is a natural
property inherited by the Voronoi-Delaunay duality. Third, and
re-emphasizing this last point, the Voronoi polygon area $A^*_h$ is
perpendicular to the hinge $h$, and provides a natural democratic
weighting for the definition of the distribution of the conical
singularity. Finally, and most importantly, locally the Regge
spacetime is an Einstein space in that the Riemann tensor is
proportional to the scalar curvature. This last property is again a
direct consequence of the relationship of the dual lattices.  The
invariance in the orientation and magnitude of the rotation bi-vector
with respect to the orientation of the dual area circumnavigated gives
this remarkable property that the lattice spacetime in RC is locally
an Einstein space \cite{E25} i.e.
\begin{equation}
R_{\mu\nu} =  \frac{R}{d}\, g_{\mu\nu}.
\end{equation}
This feature also gives the trace factor of $d(d-1)$ in front of the
Gaussian curvature of (\ref{CExp}).

Given this expression for the curvature at a hinge $h$ in a lattice
spacetime, the derivation of the RC action is
direct. Substituting the curvature expression (\ref{CExp}) and the
fundamental volume (\ref{FBlk}) into the Hilbert action (\ref{EH}).
\begin{equation}
\frac{1}{16\pi} \sum_{hinges,\ h} \left( \frac{d(d-1)}{A^*_h} \epsilon_h \right)
\left(\frac{2}{d(d-1)} A_h A^*_h \right)
\end{equation}
Canceling the obvious terms, we arrive at Tullio Regge's expression
for the Hilbert action for the lattice geometry. The Voronoi areas
cancel leaving only the simplicial (Delaunay) objects behind.  
\begin{equation}
\label{IR}
  I_R = \frac{1}{8\pi} 
\sum_{\begin{array}{c}triangle\\ hinges,\, h\end{array}} A_h \epsilon_h 
\end{equation}
While Voronoi lattice structure is absent from the Regge action, and
absent from the vacuum spacetime Regge-Einstein equations, the Voronoi
lattice is necessary for the curvature, the Einstein tensor, or the
coupling of non-gravitational sources to RC.

The RC version of Einstein's equations can now be obtained
by variation of this action with respect to the proper lengths of the
edges in the simplicial Delaunay spacetime lattice.
\begin{eqnarray}
\label{VAR}
\delta (I_R) & =  & \delta \left( \sum_{hinges,\,h} A_h \epsilon_h \right) \\
& = & \sum_{hinges,\, h} \delta (A_h) \epsilon_h + 
\underbrace{\sum_{hinges,\, h} A_h \delta (\epsilon_h)}_{zero}  = 0 
\end{eqnarray}
In the continuum the Hilbert action is varied with respect to the
metric components to yield the Einstein tensor (\ref{EH},
\ref{EH2}). However in RC we vary the Regge-Hilbert action (\ref{IR})
with respect to the edge lengths to obtain the corresponding RC
equation (\ref{reggeeq}).  The variation of the triangle area ($A_h$)
with respect to edge $L$ is expressed in terms of $L$ and the interior
angle ($\theta_h$) of triangle $h$ opposite the edge $L$.
\begin{equation}
\delta (A_h) = \frac{\partial A_h}{\partial L} \delta (L) = \frac{1}{2} L \cot (\theta_h ) 
\end{equation}
Regge showed in his original paper that the second term in (\ref{VAR})
automatically vanishes \cite{R61}.  This RC action principle yields
one equation per edge ($L_i$) in the Delaunay simplicial lattice.
\begin{equation}
\label{reggeeq}
\underbrace{
\sum_{\begin{array}{c}hinges,\ h \\ sharing\,edge\,L\end{array}}  
\frac{1}{2} L_i \cot (\theta_h ) \epsilon_h = 0
}_{
\mbox{\small Regge equation associated with edge } L_i
} 
\end{equation}

\begin{figure}
\centering
\includegraphics[width=0.5\linewidth]{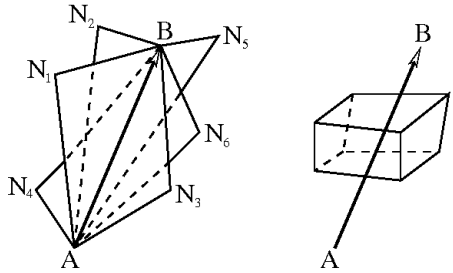}
\caption{{\it Cartain Approach to RC: }The Voronoi polyhedral cells in
  the Voronoi lattice dual to the Delaunay lattice is a natural domain
  in RC for E. Cartan's description of the Einstein tensor.  In
  particular, in a vacuum spacetime, it yields the same Regge equation
  as is obtained from the variational principle.  {\it Left:} A
  particular edge $L=AB$ in the simplicial Delaunay geometry ${\cal
    D}$ with its entourage of six triangles surrounding it. Some edges
  in ${\cal D}$ may have more or less triangles that it belongs to;
  however it must have at least four in which case the Voronoi polygon
  will be a tetrahedron. The orientation of $L$ is indicated by the
  arrow reaching from point A to point B.  The orientation of $L$ ($A$
  before $B$) induces an orientation on each of the six triangles ($A$
  before $B$ before $N_i$). {\it Right:} This orientation also
  produces an orientation on the warped cube in the Voronoi lattice
  ${\cal V}$ dual to edge $L$ in ${\cal D}$. There is a 1-1
  correspondence between any given square face on the cube and the
  corresponding triangle $A,B,N_i$.  The perimeter of each square face
  of the warped cube in $V_L$ is to be thought of as the closed loop
  encircling the corresponding triangle hinge $ABN_i$. Since the
  Voronoi lattice ${\cal S}^*$ is the circumcentric dual of ${\cal S}$
  then each vertex of the cube is located at the circumcenter of its
  corresponding simplex in ${\cal S}$.  }
\label{fig:Cartan}
\end{figure}

It is rather amazing to us that an identical equation can be obtained
directly from the lattice by using the approach first introduced by
E. Cartan (Fig.~\ref{fig:Cartan}). Recall that the Einstein tensor has two
components (\ref{EEQ}).
\begin{equation}
G=^*R^*=\underbrace{e_{\mu}}_{\begin{array}{c} Moment \\
    of \\ Rotation \end{array}}G^{\mu}_{\nu}\underbrace{dx^{\nu}}_{3-volume}
\end{equation}
In plain language, the second component of the Einstein tensor is
identified with an oriented infinitesimal 3-volume. i.e. components of
the 4-vector orthogonal to the 3-volume. The other component, Cartan
showed, can be constructed using the rotation operator, as a sum of
moments-of-rotation for each face of this 3-volume. This Cartan
approach was applied to RC by one of us (\cite{M86}) and demonstrates
the need to incorporate the Voronoi 3-volumes dual to the Delaunay
edges in order to define a precise expression for the Einstein tensor
(Fig.~\ref{fig:Cartan}). In RC the 3-volumes upon which we construct
the Einstein tensor are naturally the Voronoi polyhedrons. Given an
edge $L$ in the Delaunay lattice, one can identify its unique
polyhedron of 3-volume $V^*_L$ in the circumcentric-dual Voronoi
lattice.  The RC version of the Einstein tensor associated to edge $L$
is identical to the equation we obtained from the action
(\ref{reggeeq}) divided by the corresponding Voronoi 3-volume.
\begin{equation}
\label{creeq}
G_{LL} \, V^*_L= 
\sum_{hinges,\,h} 
\underbrace{\frac{1}{2} L \cot(\theta_h)}_{moment\, arm} 
\underbrace{\epsilon_h}_{rot'n} 
\end{equation}
In addition, the introduction of the Voronoi lattice as demanded by
the Cartan approach gives a simple geometric explanation of why there
is only one Einstein tensor component per edge in the lattice geometry
\cite{M86}. Again, the orthogonality of the Voronoi and Delaunay
lattices demand this. The moment-of-rotation vector for each face of
$V^*_L$ is oriented along edge $L$. Therefore the sum of the moment of
rotation vectors is parallel to $L$, and the orientation of Voronoi
3-volume, by definition, is parallel to $L$.  This is a direct
consequence of the mutual orthogonality that exists between each
element and its dual in the Voronoi-Delaunay lattices. We have, in RC,
a purely geometric construction of the Einstein tensor associated with
edge $L$ and it is diagonal along the edge. The RC equations requires
elements from both the Delaunay and Voronoi lattices.

The Einstein tensor is proportional to the stress-energy
tensor. Therefore in RC the flow of stress-energy ($T_{LL}$) is
directed along each edge.  This we show elsewhere leads to a
Kirchhoff-like conservation law at each vertex -- the sum of the flow
of energy-momentum into and out-of each vertex in the Delaunay lattice
must be zero. The vanishing of this flow of energy-momentum is a
consequence of a topological identity -- the boundary-of-a-boundary
is zero (\cite{W82}). 

In this section we demonstrate the indispensability of both the
Voronoi and Delaunay structures in RC. The geometric laws of
gravitation led us to couple these lattices and form a new, and
perhaps, more fundamental tiling of the spacetime with hybrid
polytopes. In the next sections to follow, we reinforce this unique
marriage between the Voronoi and Delaunay lattices and provide the
first geometric construction of the scalar curvature invariant at a
vertex (event $v$ in spacetime) that we are aware of. In the next few
sections we will use heavily the hybrid polytope and will introduce 
a reduced-version of this basic building block.

\section{The Scalar Curvature Invariant in Regge Calculus}
\label{sec5}

We have seen (\ref{EH}) that the Riemann scalar curvature invariant
plays such a central role in Einstein's geometric theory of
gravitation.  Its centrality in the theory cannot be over emphasized.
The extremum of this quantity over a proper 4-volume of spacetime,
yields a solution compatible with Einstein's field equations. It is
this scalar, so central to the Hilbert action, which yields the
conservation of energy-momentum (contracted Bianchi identities) when
variations are done with respect to the diffeomorphic degrees of
freedom of the spacetime geometry.  This scalar also augments the
Ricci tensor in coupling the non-gravitational fields and matter to
the curvature of spacetime. It not only appears in its 4-dimensional
form in the integrand Hilbert action principle of GR, it makes its
presence felt in 3-dimensions as an ``effective potential energy'' in
the ADM action.

Given this curvature invariant's pivotal role in the theory of GR, we
believe it is important to understand how to locally construct this
geometric object at a chosen event in an arbitrary curved
spacetime. Given recent interest in discrete pre-geometric models of
quantum gravity, it is ever so important to reconstruct the curvature
scalar with respect to a finite number of observers and
photons\cite{L06,AJL06}. Even though we do have familiar discrete
representations of each of the twenty components of the Riemann
curvature tensor in terms of geodesic deviation or parallel transport
around closed loops\cite{S60,B66,CD86}, and apart from the sterile act
of simply taking the trace of the Riemann tensor, we are not aware of
such a chronometric construction of the scalar curvature.

We provide such a discrete geometric description of this scalar
curvature invariant utilizing RC \cite{R61,MTW73,TW92}, and the
convergence-in-mean of RC rigorously demonstrated by Cheeger,
M\"{u}ller and Schrader \cite{CMS84}. In the spirit of quantum
mechanics and recent approaches to quantum gravity, our construction
uses only clocks and photons local to an event on an observer's world
line. Furthermore, this construction is based on a finite number of
observers (clocks) exchanging a finite amount of information via
photon ranging and yields the scalar curvature naturally expressed in
terms of Voronoi and Delaunay lattices\cite{book}.  We pointed out in
the first section of this paper that that these lattices naturally
arise in RC
\cite{CDM89,FL84,FFLR84,CFL82,CFL82b,L83,HW86,HW86b,M86,M97}.This
constriction further emphasizes the fundamental role that Voronoi and
Delaunay lattices have in the discrete representations of spacetime
which is perhaps not so surprising given its preponderant role in
self-evolving and interacting structures in nature\cite{book}. In this
analysis we re-utilize the new hybrid (half Voronoi, half Delaunay)
simplex (Fig.~\ref{fig:FB}).


The key to our derivation of the Riemann-scalar curvature is the
identification $I_R \equiv I_V$ of the usual hinge-based expression
the RC version of the Hilbert action principle \cite{R61,M97} with its
corresponding vertex-based expression. We begin with the Hilbert
action in a $d$-dimensional continuum spacetime, which is expressible
as an integral of the Riemann scalar curvature over the proper
$d$-volume of the spacetime.
\begin{equation}
I = \frac{1}{16 \pi} \int R\ dV_{proper}
\end{equation}
On our lattice spacetime, and following the standard techniques of
RC, we can approximate this action as a sum over the
triangular hinges $h$.
\begin{equation}
\label{Ih}
I \approx I_R =   \frac{1}{16 \pi} \sum_{hinges,\ h} R_h \  \Delta V_h
\end{equation}
Here, $R_h$ is the scalar curvature invariant associated to the hinge,
and $\Delta V_h$ is the proper 4-volume in the lattice spacetime
associated to the hinge $h$. Following earlier work by the
authors\cite{M97}, this curvature is defined explicitly.  Though,
non-standard in RC, we may also express the action in terms of a sum
over the vertices of the simplicial $d$-dimensional Delaunay lattice
spacetime.
\begin{equation}
\label{Iv}
I \approx  I_V = \frac{1}{16 \pi} \sum_{vertices,\ v} R_v \ \Delta V_v
\end{equation}
It is the Riemann scalar curvature ($R_v$) at the vertex $v$ that
appears in this expression that we seek in this manuscript, and it is
the equivalence between (\ref{Ih}) and (\ref{Iv}) that will yield
it. But first we must use the orthogonality inherent between the
Voronoi and Delaunay lattices to determine the relevant 4-volumes
($\Delta V_v$ and $\Delta V_h$).\footnote{Although the primary concern
  of the authors is to apply these results to the 4-dimensional
  pseudo-Riemannian geometry of spacetime, our equations are valid for
  any Riemann geometry of dimension $d$. Therefore, in the text and
  equations to follow we will explicity use the sybbol $d$ to
  represent the dimensionality of the geometry, the reader interested
  in GR can simply set $d=4$. }

\begin{figure}
\centering
\includegraphics[width=0.8\linewidth]{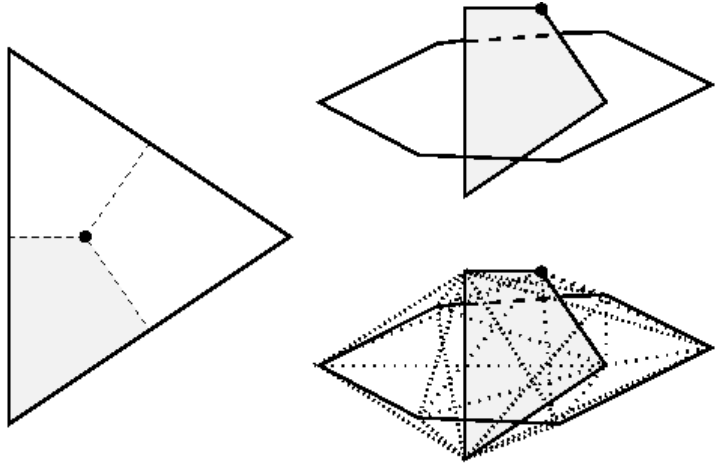}
\caption{{\it Reduced Hybrid d-cell: }The triangle hinge $h$ to the
  left is partitioned into three areas. The shaded region ($A_{hv}$)
  represents the portion of the triangle that is closer to the lower
  vertex than its other two vertices. The darkened and pronounced
  vertex appearing in each of the three line drawings of this figure
  is the circumcenter of the hinge, $h$. Each hinge has its
  corresponding 2-dimensional dual Voronoi area ($A^*_h$) shown in the
  upper right part of the figure as a pentagonal shaped polygon, and
  illustrate this dual area as ``encircling'' the $d$-dimensional
  ``kite'' hinge.  In bottom right portion of the figure, the ``kite''
  hinge is connected to its dual Voronoi polygon by ($4\times 6 = 24$)
  new lattice edges -- thus forming the reduced Voronoi-Delaunay
  hybrid d-cell which is fundamental to our derivation, the derivation
  of the Hilbert action in Regge calculus, and, we believe,
  fundamental to any discrete representation of gravitation. Each if
  these edges, as well as the edges of the Voronoi area are algebraic
  functions of the original Regge simplicial (Delaunay) lattice
  spacetime, and accordingly we have not added or subtracted any
  degrees of freedom. These new hybrid d-cells provide a new, and
  proper tiling of the lattice spacetime. }
\label{fig:ahv}
\end{figure}

Consider a vertex $v$ in the Delaunay lattice, and consider a triangle
hinge $h$ having vertex $v$ as one of its three corners. We define
$A_{hv}$ to be the fraction of the area of hinge $h$ closest to vertex
$v$ than to its other two vertices~(Fig.~{\ref{fig:ahv}}). Dual to
each triangle hinge, and in particular to triangle $h$, is a unique
co-dimension 2 area, $A^*_h$, in the Voronoi lattice. This area
necessarily lies in a $(d-2)$-dimensional hyperplane orthogonal to the
2-dimensional plane defined by the triangle $h$. The number of
vertices of the dual $(d-2)$-polygon, $h^*$, is equal to the number of
$d$-dimensional simplicies hinging on triangle $h$, and is always
greater than or equal to three.  If we join each of three vertices of
hinge $h$, with the all of vertices of $h^*$ with new edges, then we
naturally form a $d$-dimensional proper volume associated to with
vertex $v$ and hinge $h$. This $d$-dimensional polytope is a
hybridization of the Voronoi and Delaunay lattices, they completely
tile the lattice spacetime without gaps or overlaps, and they inherent
their rigidity from the underlying simplicial lattice.
\begin{equation}
\label{Vhv}
\Delta V_{hv} \equiv \frac{2}{d(d-1)} \, A_{hv}A^*_h
\end{equation}
The simplicity of this expression (the factorization of the simplicial
spacetime and its dual) is a direct consequence of the inherent
orthogonality between the Voronoi and Delaunay lattices, and its
impact on this calculation, and in RC as a whole, cannot
be overstated. These d-cells are the Regge-calculus hybrid versions of
the reduced Brillouin cells commonly found in solid state physics,
though they are hybrid because they are coupled to their dual
structures in the underlying atomic lattice. We view these as the
fundamental building blocks of lattice gravity and at the Planck scale
perhaps the RC version of Leibniz's Monads -- {\em
  Vinculum Substantiale}. The scalar factor in this expression, which
depends on the dimension of the lattice, was derived in the appendix
of an earlier paper \cite{M97}. Furthermore we obtain the complete
proper $d$-volume, $\Delta V_v$, by linearly summing (\ref{Vhv}) over
each of the triangles $h$ in the Delaunay lattice sharing vertex $v$.
\begin{equation}
\Delta V_v = \sum_{h_{|v}} \Delta V_{hv} 
                = \frac{2}{d(d-1)} \sum_{h_{|v}} A_{hv}A^*_h
\end{equation}
We now can re-express the Regge-Hilbert action at a vertex in terms of
these hybrid blocks. 
\begin{equation}
\label{Rv}
  I_V =  \frac{1}{16 \pi} \sum_{v} R_v \sum_{h_{|v}} \frac{2}{d(d-1)} A_{hv}A^*_h 
\end{equation}

\section{Obtaining the Scalar Curvature at a Vertex} 
\label{sec6}

We now return to the more familiar hinge-based Regge-Hilbert action
($I_R$) of (\ref{Ih}). The proper 4-volume associated to hinge $h$ has
been shown to be factorable in terms of the area of the triangle hinge
and its corresponding dual Voronoi area \cite{M97}.  Recall, 
\begin{equation}
\Delta  V_h = \frac{2}{d (d-1)} A_h A^*_h.
\end{equation} 
Following the procedure discussed above, we can express the area of
$h$ as a sum of its circumcentrically-partitioned pieces
(Figure~\ref{fig:ahv}).
\begin{equation}
A_h = \sum_{v_{|h}} A_{hv}
\end{equation}
Therefore the action per hinge (\ref{Ih}) can be expressed as the following
double summation:
\begin{equation}
  I_R = \frac{1}{16 \pi} \sum_h \sum_{v_{|h}} R_h \left( \frac{2}{d(d-1)} A_{hv} A^*_h\right)
\end{equation}
A key step in this derivation is the ability to switch the order of
summation, and fortunately action is unchanged if we reverse this
order.
\begin{equation}
\label{Rh}
I_R = \frac{1}{16 \pi} \sum_v \sum_{h_{|v}} R_h \left( \frac{2}{d(d-1)} A_{hv} A^*_h\right)
\end{equation} 
The vertex-based action of (\ref{Rv}) must be equal to this
hinge-based action of (\ref{Rh}). We immediately obtain the desired
expression for the Riemann scalar curvature at a vertex.
\begin{equation}
R_v = \frac{\sum_{h_{|v}} R_h A^*_h A_{hv}}
           {\sum_{h_{|v}} A^*_h A_{hv}}
= \frac{\sum_{h_{|v}} R_h A^*_h A_{hv} / \sum_{h_{|v}} A_{hv}}
       {\sum_{h_{|v}} A^*_h A_{hv}     / \sum_{h_{|v}} A_{hv}}
\end{equation}
Here we have divided the numerator and denominator by the same
quantity leaving it unchanged.  Both the numerator and denominator are
in the form of a weighted average over the "Brillion kites" ($A_{hv}$)
at vertex $v$.  We define, in a natural way, the ``kite weighted
average" at vertex $v$ of any hinge-based quantity $Q_h$ as follows:
\begin{equation}
\langle Q  \rangle_v \equiv \frac{\sum_{h_{|v}} Q_h A_{hv} }{\sum_{h_{|v}} A_{hv}}.  
\end{equation} 
Given this definition, the scalar curvature invariant at vertex $v$
can be expressed as a ``kite-weighted average'' of the integrated
hinge-based scalar curvature of RC.
\begin{equation}
R_v = \frac{\langle R_h A^*_h \rangle_v}{\langle A^*_h \rangle_v},
\end{equation}
where it was shown in \cite{M97} that the Riemann scalar curvature at
the hinge $h$ is expressible as the hinge's curvature deficit
($\epsilon_h$) per unit Voronoi area ($A^*_h$) dual to $h$.
\begin{equation}
\label{R_h}
R_h = \frac{1}{d(d-1)} \frac{\epsilon_h}{A^*_h}.
\end{equation}
Therefore the expression for the vertex-based scalar curvature
invariant derived here is strikingly similar to the usual Regge
calculus expression for the hinge-based scalar curvature invariant
(\ref{R_h}).  The only difference is that the numerator and
denominator of (\ref{R_h}) is replaced by their kite-weighted averages.
\begin{equation}
R_v = \frac{1}{d(d-1)} \frac{\langle \epsilon_h \rangle_v}
                            {\langle A^*_h      \rangle_v}
\end{equation}

\section{Convergence in Mean to the Continuum: An Example}
\label{sec7}

The importance of the Voronoi and Delaunay dual tessellations in Regge
calculus has been emphasized through their role in fundamental
quantities such as the scalar curvature.  Such vital quantities in the
discrete theory should also lead to a continuum limit in some
appropriate sense of convergence. It is not at all clear from a first
look that this measure of the scalar curvature will indeed
sufficiently approximate an appropriate smooth geometry.  Fortunately,
the result, proven earlier \cite{CMS84}, guaruntees this convergence.
Consider a triangulation of a smooth geometry, say $S^{2}$, where the
triangulation may be as fine or coarse as is desired.  As shown in
Sec.~(\ref{sec2}), the Gaussian curvature of the 2-sphere is given by
$1/R^{2}$.  Suppose, then, that instead of the smooth geometry of the
2-sphere, the geometry is given by a triangulated approximation to
$S^{2}$.  In what sense does a triangulated approximation converge to
the smooth geometry?

As we alluded to, this question was investigated in full mathematical
rigor in 1984 by Cheeger, M\"uller and Schrader \cite{CMS84}.  In
their analysis of RC's convergence to the continuum, the piecewise
flat approximation to a smooth geometry is refined into ever finer
meshes via some subdivision algorithm.  Although refinement of the
triangulation becomes finer and the edge lengths tend towards zero,
the curvature of the triangulation may not converge in a local sense,
that is, at any given vertex on the triangulation, the curvature may
vary, sometimes greatly, from the smooth geometry.  Instead, a sense
of convergence is obtained through considering extended regions of the
triangulation.  Convergence comes about in the sense of convergence in
the mean, a common theme in many examples of discrete approximations
to smooth functions.  In the case of RC, the local curvature may
fluctuate as one moves from one vertex to the other, yet given a
extended region of the triangulation, a mean curvature does converge
to the smooth geometry.  A numerical calculation makes this idea of
convergence of the discrete geometry to the smooth geometry abundantly
clear.

\begin{figure}
   \centering
   \includegraphics[width=.7\textwidth]{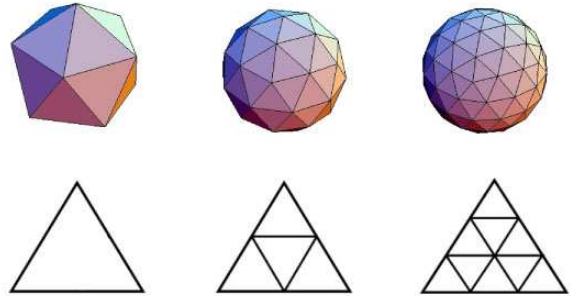}

   \caption{{\it The Geodetic Subdivision of an Icosahedron:} From
     left to right are the n=0, n=1, and n=2 frequency geodetic
     subdivisions of the icosahedron, with their face subdivisions
     shown below the triangulated 2-sphere.}
   \label{fig:S2triang}

\end{figure}

\begin{figure}
   \centering
   \includegraphics[width=.8\textwidth]{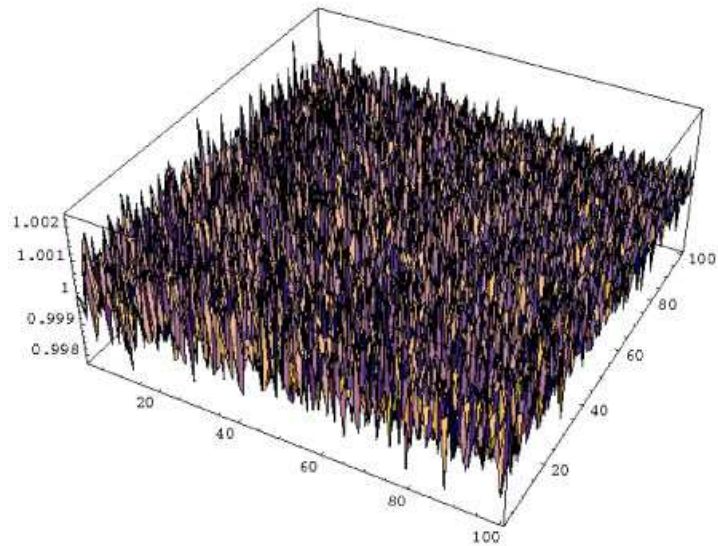}

   \caption{{\it Convergence in Mean:} For a sufficiently fine
     triangulation of the 2-sphere, the curvature at a vertex-hinge of
     the lattice may not converge to the curvature of the sphere
     locally.  This plot of the vertex curvature times the square of
     the radius of the sphere should return the continuum value of
     unity. However, across a region of an n=6 tessellated sphere,
     deviations from smooth curvature are clearly present
     ($K_vR^2=1\pm0.002$). Convergence to the smooth curvature is seen
     to be obtained in mean through considering the average curvature
     over extended regions on the surface.} \label{fig:CurvTrian}

\end{figure}

Suppose that as a first approximation to the sphere, an inscribed
icosahedron is used as the triangulation, of course one could choose
any other starting triangulation without any loss of generality. Then,
using some subdivision algorithm, the initial triangulation is
subdivided in progressively finer triangulations.
Fig.~\ref{fig:S2triang} shows a geodetic subdivision of progressively
finer triangulations of the $S^{2}$.  This geodetic subdivision is
accomplished through taking each face of the icosahedron and dividing
that face into a grid, also shown in Fig.~\ref{fig:S2triang}.  Finer
triangulations of $S^{2}$ are accomplished by taking finer meshes on
the icosahedral faces.  In order to keep the triangulations as
approximations to the smooth geometry, the new vertices are then
projected out to the surface of the sphere. The lengths assigned to
the triangulation are the geodesic lengths of their projections on
surface. In this way, a finer, and still uniform, triangulation of the
smooth spherical geometry is obtained.  In two dimensions, the
curvature is concentrated on the vertices of the triangulation, and
the Voronoi polygon associated with that vertex provides a natural
loop for parallel transport.  When the curvature is calculated for the
vertices of the triangulation of a finely triangulated geodesic dome,
one gets as expected the 1/$R^2$ curvature; however, local
fluctuations are present (Fig.~\ref{fig:CurvTrian}).  This calculation
concretely shows that the curvuture associated with the vertices of
the triangulation does not locally converge to the curvature of the
smooth geometry; however, the curvature of the triangulation does
fluctuate around the smooth geometry's curvature.  It is in this sense
that the curvature of the triangulation converges in the mean.

\section{The Fundamental Role of Voronoi and Delaunay Lattices in
  Discrete Gravity and Future Directions}
\label{sec8} 

One can't help but notice how widespread and natural the Voronoi and
Delaunay structures are in nature. The vast number of applications
described in these proceedings substantiate this as well as their
potential universality. We further reinforce this property by
demonstrating that they appear to be indispensable in providing a
discrete representation of gravity, which many would consider the most
fundamental interaction in nature. \cite{MW57} What we glean from
nature should, in the end, boils down to bit-by-bit quantum
measurements. As Wheeler would re-emphasize over and over to us; ``No
elementary quantum phenomenon is a phenomenon, unless it is brought to
close by an irreversible act of amplification.''  \cite{It-from-Bit}
These principles stand as our guide here in studying discrete models
of the fundamental forces in nature.

We analyzed here, in great depth, the inner workings of RC and its
natural expression in terms of a hybridization of the Voronoi and
Delaunay lattices. In RC the principles of GR are applied directly to
the lattice spacetime geometry. We analyzed the structure of curvature
on a simplicial lattice, the RC form of the Hilbert action of GR and
studied the structure of the Einstein tensor. Each of these central
facets of GR require the Voronoi and Delaunay lattices for their
definition.  However, what is even more surprising to us is that the
theory naturally yields a representation of spacetime by lattice cells
which inherit properties from both the Voronoi and Delaunay lattices.

Two aspects raised here may provide us with a clue as to how nature
works at the quantum level.  First, the dual hybrid lattice provides a
natural and fundamental hybrid building block describing the dynamics
of the gravitational field.  Second, the underlying discrete theory of
gravity appears more austere via the underlying factorization of the
hybrid volume elements; however, the full theory is recovered by
convergence in mean. It is these clues that we sought in this
paper. Will a unified theory of nature need to have these properties?

The hybrid block captures curvature. It promotes the Voronoi area to
``center stage'' as a natural dynamical variable in GR.  this is
hardly a new concept \cite{BRW99, M94, R93, R98, A86}, but it does
provide a new perspective to a field that has been struggling with
making a natural area-formulation of gravitation work. Anlong this
line, we ask; ``Can the Voronoi-Delaunay structure provide a
foundation to incorporate the quantum principle of complementarity in
quantum gravity''?  Perhaps the duality between the Voronoi elements and
the Delaunay elements provides a natural setting for the quantum
principle of complementarity, and perhaps the democratic partitioning
of volume supplies one with a natural topological structure, and that
dual areas are natural placeholders for the quantum Aharovov-Bohm
phases in GR , which yield the graviton fields.

\section*{Acknowledgements}We thank Arkady Kheyfets and Chris Beetle
for valuable input on this manuscript. We also thank Renata Loll and
Seth Lloyd for stimulating discussions which provided us the
motivation to continue this research, and we are especially grateful
to John A. Wheeler for providing the inspiration and initial guidance
into this field of lattice gravity, and into the foundations of the
quantum.  We applaud the conference organizers for an excellent job in
bringing together a wide spectrum of experts on Voronoi and Delaunay
tessellations in science. This conference has proved to be of great
value in our own field of research.  We would like to thank Florida
Atlantic University's Office of Research and the Charles E. Schmidt
College of Science for the partial support of this research.


\begin{thebibliography}{99.}


\bibitem{R61} T Regge: {\it Nuovo Cimento} {\bf 19}, 558 (1961)

\bibitem{TW92} P.A. Tuckey, R.M. Williams: {\it Class. Quantum
    Grav.} {\bf 9}, 1409-1422 (1992)

\bibitem{book} A. Okabe, B. Boots, K. Suqihara, S. N. Chiu: {\it
    Spatial Tessellations: Concepts and Applications of Voronoi
    Diagrams} 2nd ed. (John Wiley \& Sons Ltd., West Sussex 2000)


\bibitem{CMS82} J. Cheeger, W. M\" uller, R. Schrader: {\it
    Lecture Notes in Phys.} {\bf 160}, 176-188 (1982)

\bibitem{CMS84}  J. Cheeger, W. M\" uller, R. Schrader: {\it
    Commun. Math. Phys.} {\bf 92}, 405 (1984)

\bibitem{H86} H. Hamber: Simplicial Quantum Gravity. In: {\it Gauge
    Theories, Critical Phenomena and Random Systems}:
    Proceedings of the 1984 Les Houches Summer School, Session XLIII,
    ed by K.  Osterwalder, R. Stora (North Holland, Amsterdam 1986)

\bibitem{M86} W.A. Miller: {\it Found. Phys.} {\bf 16}, 143-169 (1986)

\bibitem{M92} P.A. Morse: {\it Class. Quantum Grav.} {\bf 9}, 2489-2504 (1992)

\bibitem{L82} T.D. Lee: {\it Phys. Lett.} {\bf B122}, 217-220 (1982)

%

\bibitem{MTW73} C.W. Misner, K.S. Thorne, J.A. Wheeler: {\it Gravitation}
(W. H. Freeman and Company, New York 1973) ch~42

\bibitem{E25} L.P. Eisenhart {\it Riemannian Geometry} (Princeton
  Univ. Press, Princeton 1925) Ch.~2

\bibitem{W82} J.A. Wheeler: Particles and geometry. In {\it
    Unified Theories of Elementary Particles}, (Springer-Verlag, San
  Francisco 1982)  ed by Breithenlohner, H. Durr pp.~189-217

\bibitem{L06} S. Lloyd: ``A theory of quantum gravity based on
  quantum computation'' {\it arXiv:quant-ph/0501135} (2006)

\bibitem{AJL06} J. Ambjorn, J. Jurkiewicz, R. Loll: ``Quantum
  gravity: the art of building spacetime'' {\it arXiv:hep-th/0604212v1
  } (2006)

\bibitem{S60} L. Synge: {\it Relativity, The General Theory}
  (North Holland, Amsterdam 1960) p~408

\bibitem{B66} B. Bertotti: {\it J. Math. Phys.} {\bf 7}, 1349 (1966)

\bibitem{CD86} I. Ciufolini, M. Demianski: {\it Phys. Rev.} {\bf
    D34}, 1018-1020 (1986)



\bibitem{CDM89} M. Castelle, A. D'Adda, L. Magnea {\it Phys. Lett}
  {\bf B232}, 457 (1989)

\bibitem{FL84} R. Friedberg, T.D. Lee: {\it Nucl. Phys.} {\bf
    B242} p~145 (1984)

\bibitem{FFLR84} G. Feinberg, R. Friedberg, T.D. Lee,  H. C. Ren: {\it
    Nucl. Phys.} {\bf B245}, 343 (1984)

\bibitem{CFL82} N.H. Christ, R. Friedberg, T.D. Lee: {\it
    Nucl. Phys.} {\bf B202}, 89 (1982)

\bibitem{CFL82b} N.H. Christ, R. Friedberg, T.D. Lee: {\it
    Nucl. Phys.} {\bf B210}, 337 (1982)

\bibitem{L83} T.D. Lee, {\it Phys. Lett.} {\bf B122}, 217 (1983)

\bibitem{HW86} H. Hamber,  R.M.  Williams: {\it Nucl. Phys.} {\bf
    B267}, 482 (1986)

\bibitem{HW86b} H Hamber, R. M.  Williams: {\it Nucl. Phys.} {\bf
    B248}, 392 (1986)

\bibitem{M97} W.A. Miller: {\it Class. Quantum Grav.} {\bf 14}, L199-L204 (1997)

\bibitem{MW57} C. W. Misner, J. A. Wheeler, {\it Ann. Phys.} {\bf 2}, 525-603 (1957)

\bibitem{It-from-Bit} J. A. Wheeler, Information, Physics, Quantum:
  The Search for Links. in: {\it Complexity, Entropy, and the Physics
    of Information,} ed. by W. H. Zurek (Addison-Wesley, Redwood City,
  CA 1990) pp. 3-28.  (1990).

\bibitem{BRW99} J. W. Barrett, M. Rocek, R. M. Williams, {\it Class. Quantum Grav.} {\bf 16}, 1373-1376 (1999)

\bibitem{M94} J. M\" akel\" a, {\it Phys. Rev.} {\bf D49}, 2882 (1994)

\bibitem{R93} C. Rovelli, {\it Phys. Rev.} {\bf D48}, 2702 (1993)

\bibitem{R98} C. Rovelli, {\it Living Rev. Rel.} {\bf 1}, 1 (1998)

\bibitem{A86} A. Ashtekar, {\it Phys. Rev. D} {\bf 36}, 1587-1602 (1986)


	

\end{thebibliography}
\end{document}